\documentclass[doublecol]{epl2} 
  \usepackage{color}
  \usepackage{newlfont}
  \usepackage{graphicx}
  \usepackage{amssymb}
  \usepackage{amsmath}
  \usepackage[latin1]{inputenc}
  
\newcommand{\be}{\begin{eqnarray}}
\newcommand{\ee}{\end{eqnarray}}

\title{Model for the resistance force acting on circular bodies in the imminence of rolling}
\shorttitle{Model for the resistance force acting on circular bodies} 

\author{A. L. O. Bilobran\inst{1} \and R. M. Angelo\inst{1}}
\shortauthor{A. L. O. Bilobran \etal}

\institute{                    
  \inst{1} Departamento de Física, Universidade Federal do Paraná, Caixa Postal 19044, 81531-980, Curitiba, PR, Brazil
}
\pacs{62.20.Qp}{Friction, tribology, and hardness}
\pacs{46.55.+d}{Tribology and mechanical contacts}
\pacs{45.40.-f}{Dynamics and kinematics of rigid bodies}

\abstract{The laws of friction are reasonably well understood for the case of blocks in contact with rough plane surfaces. However, as far as bodies with circular sections are concerned, the physics of friction becomes more involving and it is not possible to adopt a simple conceptual framework to explain all phenomena. In particular, there is no approach so far to the problem of the resistance force that opposes to circular bodies that remain at rest while acted upon by small forces. Here we fill this gap by introducing a mechanical model based on both the elasticity theory and Hertz contact mechanics. Our approach furnishes a quantitative expression for the critical force beyond which rest can no longer be maintained. Besides confirming the expected proportionality of the resistance force with the load, our result contains no free parameters and is expressed solely in terms of physical properties of the problem, such as the pressure of the body per unit of superficial area, a relation between the Young modulus of the surface and its Poisson ratio, and the symmetry of the contact.}			
			
\begin{document}

\maketitle

\section{Introduction}

When two solid bodies in contact move relatively to one another (or when a resting body is acted upon by a tiny applied force) there appears a resistance force, contrary to the (onset of) motion, which we generically call friction. Estimates point out that up to $6\%$ of the GNP of a country may be wasted with energy losses associated with friction~\cite{persson00}. Such a contact interaction permeates a number of aspects of human life, being important even for everyday acts such as walking and wearing. Despite the undisputed relevance of friction and its longterm history in science, it is fair to say that its basic physics is not yet fully understood.

The first attempts to describe the physics of friction date back to Leonardo da Vinci, who empirically derived the basic laws of friction. In his {\em Codex-Madrid I} (1495), da Vinci summarized that friction is i)~proportional to the load and ii)~independent of the contact area of the sliding surfaces. Also, he was the first to introduce the term {\em coefficient of friction} and to experimentally determine its typical value of $1/4$. These results were rediscovered in 1699 by the French physicist Guillaume Amontons. Later on, others contributed to the foundations of the science of Tribology. In 1750, Leonard Euler proposed that friction was caused by interlocking irregularities between the surfaces. Also, he introduced the symbol $\mu$ for the coefficient of friction and was the first to make a distinction between static and kinetic friction. It is due to Charles Augustin Coulomb (1781) both the result according to which there exists a maximum static friction proportional to the load and the third classic law of friction, which states that iii)~kinetic friction is independent of the sliding velocity. It was not until the middle of the last century, that the laws of friction start to receive other than purely empirical support~\cite{leben39}. In 1950, Frank Philip Bowden and David Tabor~\cite{tabor} proposed a fundamental physical explanation for friction. According to their approach, the contact between surfaces is due to asperities and, as a consequence, the actual contact area, which friction ultimately depends on, is small compared to the apparent area. Thus, the larger the applied load the larger the real contact area, which in turn implies larger friction.

As far as bodies with circular sections (mainly cylinders and spheres) are concerned, the physics of friction becomes more subtle. It has been recognized since long ago that in situations involving the so called {\em rolling friction} it is no longer possible to neglect the deformations of the bodies~\cite{reynolds}. In fact, studies on this subject often appear connected with the Contact Mechanics Theory~\cite{popov,ll,persson}, whose founder father is Heinrich Hertz. Although his formulation neglects adhesion and Van der Waals interactions, the contact area can often be analytically derived in terms of the physical properties of the bodies. The adhesive interaction and the Van der Waals interaction are sequentially added to Hertz's approach via the JKR and DMT theories, respectively~\cite{johnson85}. Tabor demonstrated that both theories are successful, each one in a specific regime, which is defined by the materials and geometry involved. The gap between those two is filled by the Maugis-Dugdale theory. The JKR and DMT theories attribute the energy loss to hysteresis instead of slip, which explains why it is barely affected by lubricants~\cite{tabor61}. The reader is referred to refs.~\cite{greenwood97,zhao} for a survey of all the aforementioned theories and their regimes of validity. In many treatments, rolling friction appears as a torque $\rho P$ opposed to the rotation of the body, where $P$ is the load and $\rho$ is the coefficient of rolling friction. The literature points out that $\rho$ is in the range $10^{-2}$ to $10^{-3}$ cm~\cite{eensign}, although some discordance exists~\cite{nelson}.

The problem of rolling friction was investigated in a variety of interesting situations, most of them involving macroscopic regimes~\cite{atack,bueche,norman,cebrian,poschel,brilliantov,ertas99,wolf03,wolf05}. Recently, however, new insights have been put forward by adopting a microscopic perspective (see refs.~\cite{braun10,yu03} for a review on this topic), by means of which the origins of friction are explored in the atomic domain. Efforts in this direction, often conjugated to friction- and atomic-force microscopy~\cite{welland97,wiesen97,johnson97}, define the area referred to as {\em Nanotribology}~\cite{wiesen95,lu99,reimann12}. In this context, the problem of friction between two substrates separated by a thin lubricant film was theoretically addressed via molecular dynamics simulations in simple two-dimensional models involving Lennard-Jones potentials~\cite{braun05}. Experimentally, setups based on atomic force microscopy have been used (i) to directly measure the adhesion and rolling-friction forces between silica microspheres~\cite{butt99} and (ii) to measure and control the rolling and sliding of carbon nanotubes at nanometric scale~\cite{superfine99}.

This paper aims to contribute to an issue which, to the best of our knowledge, remains open so far. It refers to a very basic problem with rather nontrivial solution. We are concerned with the resistance force that acts on a circular macroscopic body (sphere or cylinder) while the static equilibrium is maintained, i.e., while the body remains at rest. This is analogous to the situation in which a block remains at rest on a rough horizontal surface while acted upon by a tiny force $\mathbf{F}=F\mathbf{e}_x$; a frictional force $\mathbf{f}=-F\mathbf{e}_x$ appears in the contact area which prevents the motion from occurring. This certainly cannot be the whole story for circular bodies. Indeed, it is easy to see that if the external force is applied horizontally, e.g., towards the center of mass of the body, those forces would produce a resulting torque in relation to center of mass and the static equilibrium would no longer exist. Nevertheless, we do know from our everyday experience\footnote{When we breath in front of a bowling ball the air pressure is not sufficient to breakdown the static equilibrium. Actually, we can significantly increase the pressure before the ball starts to move.} that static equilibrium must exist for a sufficiently small $F$. Our approach consists of using results of the Hertz Contact Mechanics to build, out of a minimum set of assumptions, a mechanical model for the resistance force acting on macroscopic circular bodies. We derive an upper bound for the maximum force that can be applied before the body starts to move. Free from any fitting parameter, our formula is shown to depend only on physical quantities, such as the load, the geometry of the body, and the elastic properties of the surface.

\section{Model}

Our aim is to devise a mechanical model without any appeal for microscopic details of matter, such as its atomic structure. As mentioned above, we are concerned with the forces that act while the static equilibrium is preserved. Moreover, we focus on cases in which the static regime is necessarily replaced by rolling without sliding. For concreteness, we can imagine the onset of macroscopic motion of a bowling ball (initially at rest on a horizontal surface) when acted upon by a force that smoothly increases from zero.

\subsection{Contact force}
\begin{figure}[t]
\centerline{\includegraphics[scale=0.17]{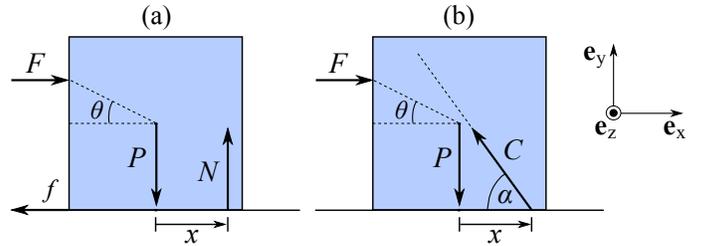}}
\caption{(a) An external force $\mathbf{F}=F\mathbf{e}_x$, smaller than the maximum static friction, is applied on a cubic block of edge $2l$ and weight $\mathbf{P}=-P\mathbf{e}_y$ initially at rest. Contact forces $\mathbf{f}=-f\mathbf{e}_x$ and $\mathbf{N}=N\mathbf{e}_y$ maintain the cube at rest. The rotational static equilibrium is preserved when $x=\tfrac{F}{P}l(1+\tan\theta)$. (b) The two contact forces are merged into a single contact force $\mathbf{C}$, whose inclination is given by $\tan\alpha=P/F$.} 
\label{fig1}
\end{figure}

Let us start by considering the problem of a cubic block of weight $\mathbf{P}=-P\mathbf{e}_y$ and edge $2l$ acted upon by a force $\mathbf{F}=F\mathbf{e}_x$ whose modulus is smaller than the maximum static friction $\mu N$. The situation is depicted in fig.~\ref{fig1}. As usual, to maintain the {\em translational} static equilibrium one should demand the contact forces to be given by $\mathbf{f}=-f\mathbf{e}_x$ (friction) and $\mathbf{N}=N\mathbf{e}_y$ (normal), with $f=F$ and $N=P$. 

Now, in order to preserve the {\em rotational} static equilibrium, the normal force has to act a distance $x=\frac{F}{P} l(1+\tan\theta)$ far from the center of symmetry (see fig.~\ref{fig1}(a)), a result that immediately follows from the basic laws of classical mechanics under the assumption that the total torque vanishes. We see that the point of application of the normal force moves sideways as either $F$ or $\theta$ increases. In practice, this effect may derive from unperceivable rotations and deformations that make the microscopic contact area move laterally. When $F(1+\tan\theta)>P$, one has that $x>l$, implying that the rotational equilibrium can no longer be maintained. The point of application of the friction force $\mathbf{f}$ does not influence the torque in relation to center of mass, but presumably it should be where the contact is more effective. One may naturally expect that both $\mathbf{f}$ and $\mathbf{N}$ effectively act on the same region. 

This observation makes it possible to merge the friction and the normal forces into a {\em single contact force} $\mathbf{C}$ [see fig.~\ref{fig1}(b)], from which we may derive $\mathbf{f}=(\mathbf{e}_x\cdot\mathbf{C})\,\mathbf{e}_x$ and $\mathbf{N}=(\mathbf{e}_y\cdot\mathbf{C})\,\mathbf{e}_y$. Accordingly, the static equilibrium will occur for $\mathbf{C}+\mathbf{F}+\mathbf{P}=\mathbf{0}$ and $x=\frac{F}{P}l(1+\tan\theta)$. Of course, we are adopting an idealization of a single point of application for the contact force, which actually is a resulting force that encapsulates the global reaction of the surface. This approach is particularly interesting for problems involving circular bodies, for which the effective contact area is rather smaller.

\subsection{Rigid bodies and elastic surfaces}

\begin{figure}[t]
\vspace{0.2cm}
\centerline{\includegraphics[scale=0.2]{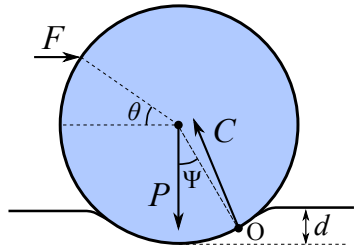}}
\caption{A rigid circular body of radius $R$ and weight $\mathbf{P}$ deforms an elastic plane up to a maximum penetration length $d$. Under the application of a tiny force $\mathbf{F}$ a reaction force $\mathbf{C}$ emerges which guarantees the static equilibrium. The angular position of the point O at which $\mathbf{C}$ applies is $\Psi$.}
\label{fig2}
\end{figure}

We focus on regimes in which an initially plane surface deforms elastically under the  weight $P$ of an ideally rigid circular body of radius $R$. The situation is illustrated in fig.~\ref{fig2}. When the body is acted upon by a sufficiently small external force $\mathbf{F}$, a contact force $\mathbf{C}$ appears in the point of application O in order to preserve the static equilibrium. The angular displacement of O from the vertical is denoted by $\Psi$ and the {\em penetration length}  by $d$. From the geometry of fig.~\ref{fig2} it is straightforward to check that $\Psi \in [0,\Psi_{g}]$, where
\be
\Psi_{g}\cong \left(\frac{2d}{R}\right)^{1/2},
\label{Psi_g}
\ee
``$g$'' standing for ``geometric.''
The quality of this approximation is determined by the elastic properties of the surface, in particular by its {\em Young's modulus}, $E$. The larger the value of $E$ the smaller the deformation $d$ and the better the approximation \eqref{Psi_g}. Note that the very notion of $\Psi_g$ is not much accurate because the actual contact area is smaller than that suggest by the geometry. In other words, because of the smoothness of the surface---there should not be any discontinuity in the first derivative at any point of the surface---the last point of contact actually occurs at an angle slightly smaller than the geometrical value given by eq.~\eqref{Psi_g}.

An important advantage of the present framework is that the Contact Mechanics Theory furnishes several analytical results describing the characteristics of the elastic deformation in terms of the physical properties of the bodies involved~\cite{popov,ll}. A fundamental building block of this theory is the formulation of the correct pressure distribution. Theoretical treatments of these problems can be found in refs.~\cite{popov,ll,johnson85,formulas}. Of particular interest for our purposes are the formulas of the penetration length for a {\em sphere} of radius $R$, 
\begin{subequations}
\be
d_{\text{sphere}}=\left(\frac{9}{16}\frac{P^2}{\kappa^2 R}\right)^{1/3},
\label{dS}
\ee
and a {\em cylinder} of radius $R$ and length $L$,
\be
d_{\text{cylinder}}=\frac{4}{\pi} \frac{P}{\kappa  L}\gamma.
\label{dC}
\ee
\label{d}
\end{subequations}
As usual, the results are written in terms of the parameter $\kappa =E/(1-\nu^2)$, here referred to as {\em medium constant}, as it gives a relation between the Young modulus $E$ and the Poisson ratio $\nu$ of the medium. Although the result~\eqref{dS} is universally accepted, for the cylinder there is still some discordance. Indeed, by direct inspection one may verify that $\gamma=\tfrac{3}{8}\ln\big(\frac{2\pi e}{3}\frac{\kappa L^3}{PR}\big)$ in ref.~\cite{formulas} whereas $\gamma=1$ in ref.~\cite{popov}, the former enjoying some experimental support~\cite{thwaite69}. Note that formulas~\eqref{d} accurately hold only in the Hertz regime, i.e., when the adhesion forces can be neglected in comparison with the weight of the body. Off this regime, the above formulas should be superseded via appropriate theories~\cite{greenwood97}.

\subsection{Maximum resistance force}

For blocks, while the static equilibrium is maintained under the action of an external force, the horizontal reaction of the surface is generically called static friction. For circular bodies there is some ambiguity, for some authors associate the term with the situation in which the body rolls without sliding. To avoid confusion, here we call {\em resistance force} the total horizontal reaction offered by the contact force. Formally, however, we preserve the notation
\be
\mathbf{f}\equiv(\mathbf{e}_x\cdot\mathbf{C})\,\mathbf{e}_x,
\label{f}
\ee
which obviously leads to $\mathbf{f}=-F\mathbf{e}_x$ in the present problem. It is immediately seen that the resistance force is null for $F=0$, as in this case $\mathbf{C}=P\mathbf{e}_y$. If we fix the modulus $F$ of the external force and increase the angle of application $\theta$, then $\Psi$ should increase to balance the torque, while the modulus of $\mathbf{C}$ keeps fixed to guarantee the translational static equilibrium (see fig.~\ref{fig2}). On the other hand, if we fix $\theta$ and increase $F$, then both the modulus and the inclination $\alpha$ (defined as in fig.~\ref{fig1}) of the contact force in relation to the $x$ axis has to vary in order to ensure that $\mathbf{C}+\mathbf{F}+\mathbf{P}=\mathbf{0}$. In this case as well, $\Psi$ has to increase to preserve static equilibrium. Clearly, $\Psi$ is a function of the applied torque, i.e., $\Psi=\Psi(F,\theta)$. For simplicity, in what follows we will associate the notion of {\em maximum resistance force} specifically with $\theta=0$.

From the condition that the total torque keeps null in relation to O, it is not difficult to prove that $F=P\tan\Psi$. This relation confirms that $\Psi$ increases with $F$. It then follows from \eqref{f} that $f=||\mathbf{f}||=\mu\,P$, with $\mu\equiv \tan\Psi$, where we have introduced $\mu$ to make the link with the theory for blocks. Now we look for the maximum value of $\Psi$ that ensures static equilibrium. Presumably, this value depends on the physical properties of the surface, but it must not be greater than the geometrical bound $\Psi_g$ given by eq.~\eqref{Psi_g}, otherwise the point of application of the contact force would be out of the contact region. Our model concerns situations in which no sliding occurs, so that we assume that the rolling will start at some value $\Psi$ for which the surface sinks under the pressure caused by the circular body. The perturbation induced by the critical force on the surface tends to propagate initially as a wave. From this stage on, theories of rolling friction can be invoked to explain the motion~\cite{reynolds,popov,johnson85}, but this is out of the scope of this contribution. Given that $\Psi\leqslant \Psi_{g}$, the maximum resistance force the surface can provide before the rolling starts is upper bounded as
\be
f_{\max}\leqslant\mu\,P, \qquad \text{with} \qquad \mu=\tan\Psi_{g}.
\label{fmax}
\ee
This relation [along with eqs.~\eqref{Psi_g} and \eqref{d}] defines our model for the maximum resistance force. Although the above relation provides just an upper bound, it furnishes a quantitative formula with no free parameter. In addition, our result depends only on physical properties that are generally known, so that the adequacy of model can be experimentally assessed.

\subsection{Unified view}

Another remarkable aspect emerges from our simple model, namely, the possibility of unifying the results for spheres and cylinders in a single formula. Consider the superficial areas $S=4\pi R^2$ (sphere) and $S=2\pi RL$ (cylinder). Define, for future convenience, the following dimensionless quantity:
\be
\wp \equiv \frac{P/S}{\kappa}=\frac{\text{superficial pressure}}{\text{medium constant}}.
\label{wp}
\ee
The pressure ratio $\wp$ is clearly a quantity that encapsulates information about both participants of the phenomenon, the indenter (the circular body) and the medium (the elastic surface). Roughly, it measures the local capability of the body in to deform the surface.

Now, since we are focusing in regimes in which the penetration length $d$ is small, one may use Eqs.~\eqref{Psi_g} and \eqref{fmax} to approximate the coefficient of resistance as $\mu\cong\Psi_{g}\cong (2d/R)^{1/2}$. From Eqs.~\eqref{d} and \eqref{wp} it is straightforward to show that one may unify the results for the sphere and the cylinder into
\be
\mu\cong\,\beta_D\,\,\wp^{1/(D+1)}, 
\label{mu}
\ee
where $\beta_D\equiv(5-D)\gamma^{(2-D)/2}$. The result for the sphere (cylinder) is retrieved with $D=2$ ($D=1$). Interestingly, the constant $D$ can be directly associated with the symmetry of the contact: While for the sphere the distribution of pressure is two-dimensional, for the cylinder it is one-dimensional~\cite{popov,johnson85}. It is instructive to imagine the situation right after the beginning of the contact, when the body has just been released on the surface. For the sphere, the contour lines of pressure are described by concentric circumferences (say, symmetry $D=2$) while for the cylinder they form concentric thin long stripes (symmetry $D=1$). This observation further valorizes the result \eqref{mu}, as it encodes into the resistance coefficient information about the symmetry of the contact. 

\begin{figure}[t]
\vspace{0.3cm}
\centerline{\includegraphics[scale=0.35]{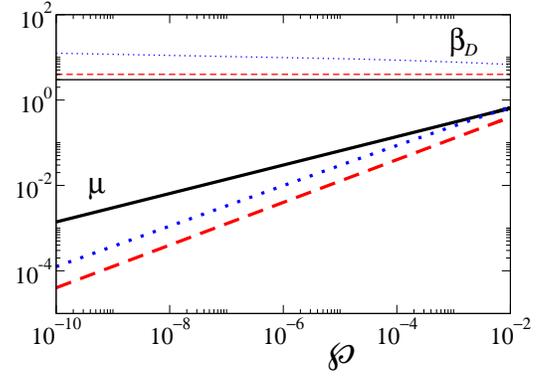}}
\caption{Log-log plot for both $\mu$ (thick lines) and $\beta_D$ (thin lines) as a function of $\wp=P/\kappa S$ [see eq.~\eqref{mu}]. The color code is as follows: solid black lines for $\beta_2$ (sphere), dashed red lines for $\beta_1^{(1)}$ (cylinder, ref.~\cite{popov}), and dotted blue lines for $\beta_1^{(2)}$ (cylinder, ref.~\cite{formulas}). In these simulations, $L=5R$.}
\label{fig3}
\end{figure}

As far as the parameter $\beta_D$ is concerned, some comments are in order. While for the sphere $\beta_2=3$, for the cylinder on has that either $\beta_1^{(1)}=4$, in accordance with ref.~\cite{popov}, or $\beta_1^{(2)}=\big[6\ln\big(\tfrac{eL^2}{3\wp R^2}\big)\big]^{1/2}$ (see ref.~\cite{formulas}), the latter keeping an explicit dependence on $\wp$. Because of its logarithmic form, $\beta_1^{(2)}$ applies as a meaningful physical parameter only if $\wp\lesssim (L/R)^2$. While this condition seems to further restrict the applicability of the model, it actually is in full harmony with the premises of our approach. To check that, one should note, on one hand, that the formula \eqref{dC} was derived in ref.~\cite{formulas} under the assumption that $L\gg R$ and, on the other hand, that we have previously required that $\wp \ll 1$, which corresponds to the regime of hard elastic media and hence small penetration lengths. It follows, therefore, that $\beta_1^{(2)}$ is indeed a well behaved parameter in the range $\wp\ll (L/R)^2$ within which our model applies. Furthermore, likewise $\beta_2$ and $\beta_1^{(1)}$, $\beta_1^{(2)}$ does not significantly vary throughout the (in principle permissible) several decades for $\wp$ (see fig.~\ref{fig3}), so that the unified formulation $\mu\propto \wp^{1/(D+1)}$ reveals to be meaningful indeed.

Our results were derived for a plane surface, negligible adhesion forces, and smooth contact areas. Although it is not clear whether our final formula for $\mu$ will preserve its features off these regimes, the logical structure of our model, based on the mechanical equilibrium of forces and torques, should still apply. That is because, within the rigid-body approximation, the geometry of the contact with the elastic substrate will remain the same. Of course, adaptation of some ingredients will be necessary. For instance, for a curved surface one needs to replace eqs.~\eqref{d} with their pertinent substitutes~\cite{popov,formulas}, while eqs.~\eqref{Psi_g} and \eqref{fmax} will stay the same. Adhesion forces and roughness might be introduced approximately in the model by taking the penetration length predicted by the JKR theory~\cite{popov,johnson85} and adopting an effective weight $P+F_{\text{ad}}$ for the circular body. In such a formulation, the resulting adhesive force $F_{\text{ad}}$ should encompass the fact that the maximum resistance force is expected to increase with adhesion and decrease with the roughness of the contact~\cite{popov,zhao,yu03}. The applicability of our framework is nevertheless conditioned to the regime in which the characteristic length of the asperities is much smaller than the radius of the body under concern.

\section{Concluding remarks}
The physics of the friction is widely addressed via phenomenological approaches. The formula $\mu P$ is universally accepted for both the static and the kinetic friction, with $\mu$ encapsulating all the unknown physics. The situation is not different when the problem of rolling friction is concerned. In this paper, we have modeled the horizontal force a surface applies to a circular body when the latter remains at rest while acted upon by an external force. To some extent, this resistance force stands for the problem of circular bodies as the static friction stands for blocks. To the best of our knowledge, this problem has never been addressed so far.

Our approach consists of a minimum mechanical model for the contact of an ideally rigid circular body, such as a cylinder or a sphere, with an elastic plane surface of known properties. By assuming that both the resistance and the normal forces are components of a single contact force $\mathbf{C}$ we have derived an upper bound for the maximum resistance force that the surface can provide before the rolling starts. Our main result---composed of eqs.~\eqref{fmax} and \eqref{mu}---shows that the maximum resistance force is proportional to the load $P$ via a coefficient that encapsulates all the physics of the contact. Indeed, $\mu$ is shown to depend solely on physical properties of the problem, namely, the pressure per unit area caused by the circular body, the medium constant, defined by a relation between the Young modulus and the Poisson ratio, and the symmetry of the contact. In particular, the ratio $\wp$ between the superficial pressure ($P/S$) and the medium constant in units of pressure ($\kappa$), emerges as distinctive figure of merit for the physics of $\mu$. 

We hope our results can motivate further explorations on this basic problem. Experimentally, the model could be readily tested in the macroscopic domain. On the theoretical side, an exciting research program would involve the study of the effects of adhesion and Van der Waals forces, the roughness of the contact, and the elasticity of the circular body. With such improvements, we believe our model could also be assessed, with current technologies, in the context of controlled manipulation at microscopic scale~\cite{hashimoto00,sitti07}.

\acknowledgments
The authors gratefully acknowledge A. F. D. Modtkowski and A. D. Ribeiro for insightful discussions and M. Evstigneev for valuable comments on an early version of this paper.


\end{document}